\documentclass[twocolumn,showpacs,preprintnumbers,amsmath,amssymb,A4paper]{revtex4}
\usepackage{graphicx}% Include figure files
\usepackage{dcolumn}% Align table columns on decimal point
\usepackage{bm}% bold math

\begin{document}
\newcommand{\kp}{{\bf k$\cdot$p}\ }
\def\blm#1{\textit{\textbf{#1}}}
%\preprint{APS}

\title{Optical pumping of the electron spin polarization in bulk CuCl}
%\\with Forced Linebreak}% Force line breaks with \\

\author{W.Ungier, R.Buczko}
 \affiliation{Institute of Physics, Polish Academy of Sciences\\
       Al.Lotnikow 32/46, 02--668 Warsaw, Poland\\
 %  e-mail: buczko@ifpan.edu.pl
  }

\date{\today}

\begin{abstract}
In CuCl bulk crystal negatively charged excitons (trions $X^-$)
can be induced by the resonant optical excitation of extra
electrons in conduction band minimum. In the case of circularly
polarized light and due to the top valence band structure only the
electrons with one spin orientation (for example antiparallel to
the direction of the light propagation in the case of $\sigma ^+$
polarization) contribute to the formation of trions $X^-$. At the
same time spontaneous decay of $X^-$ populates both, parallel and
antiparallel electron spin states, thus producing an enhancement
of one spin orientation. We propose to use this mechanism for
optical pumping of electronic spin polarization. We estimate the
momentum matrix elements of electron - trion transitions which are
the main factors determining pumping rate. The pumping dynamics is
described by the density matrix evolution equations which couple
to the Maxwell wave equation for the coherent pumping laser pulse
propagating through the sample. The results of our simple model
calculations suggest that spin polarization close to 100\% can be
achieved in CuCl in the time scale of the order of 100ps.
\end{abstract}

% Classification Scheme
%Use showkeys class option if keyword

\maketitle

\section{\label{sec:level1}Introduction}

In many semiconductors the spin lifetime of an electron at the
bottom of the conduction band is much longer than other relevant
time scales. Therefore the electron spin, as a carrier of
information, has big potential in many possible spintronics and
quantum informatics applications. A basic requirement for the
workable devices is the possibility of sufficient initial electron
spin polarization. One of the known methods is the injection of
the electrons into semiconductor media through the spin aligner
(magnetic or semimagnetic contact)
\cite{Ostreich,Fiederling,Ohno,Jonker,Ghali}. Recent achievements
of this technique claim 85\% efficiency
\cite{Ostreich,Fiederling}. Another way is the optical excitation
of semiconductor media with circularly polarized light which
creates spin polarized electrons and holes \cite{Zutic,Dyakonov}.
For example, in bulk GaAs, the representative zinc-blende direct
gap semiconductor, the ratio of densities of excited electrons
with opposite spin polarizations reaches 3:1 due to the optical
selection rules for the transitions from heavy and light hole
subbands. In low-dimensional structures, because of the split-off
of the heavy and light hole subbands, the degree of electron spin
polarization can in principle reach 100\%. However, the spin
polarization of electrons excited from the valence bands is
transient. The time of its persistence is limited mainly by the
lifetime of electrons which can recombine with holes. The
electrons and holes separation by an electric field is needed in
order to extend the spin lifetime.

In the case of single electrons in quantum dots (QDs) many
possible ways of spin initialization and manipulation have been
proposed. Among them several interesting proposals refer to
optical control
\cite{Imamoglu,Anisimovas,Shabaev,Pochung,Gywat,Bracker,Dutt}. The
electron in QD can be resonantly excited into the heavy hole trion
$X^-$ state \cite{Glasberg}. Shabayev et al. \cite{Shabaev}
suggest to use the trion as an intermediate state for feasible
initialization of the electronic spin. In order to obtain the well
defined spin polarization, regardless of its initial state, they
propose to use circularly polarized optical $\pi$ pulse combined
with $\pi$ pulse of transverse magnetic field and successive
spontaneous trion decay.

One can think about using the same or similar mechanism as
proposed for QDs for electronic spin polarization in the case of
small concentration of excess free electrons in semiconducting
quantum well structure or bulk material. However, the binding
energy of trion $E_{b2}$ (i.e. binding of the second electron to
the exciton) in bulk is usually much smaller than 1meV and even in
2dim quantum wells it rarely exceeds 1meV
\cite{Glasberg,Kheng,Finkelstein,Shields,Hayne,Esser}, making the
selective creation of trions rather difficult. Furthermore the top
of the valance band is fourfold degenerate in the typical
zinc-blende semiconductors and due to the selection rules for the
trion creation it would be rather hard (but not impossible
providing the trion's spin relaxation time is long enough in
comparison to the trion lifetime) to reach high electronic spin
polarization in the bulk material. Nevertheless in the special
case of copper halides, because of the very small ratio of
electron to hole effective masses $\sigma=m_e/m_h$ in these
materials, the binding energy of $X^-$ can appear to be relatively
large. In CuCl, which is widely used for direct creation of
biexcitons and recently also for generation of entangled photons
\cite{Edamatsu} the effective mass of the hole is high
($m_h\approx 20 m$, $m_e\approx 0.4 m$), and because of that
$E_{b2}$ is expected to be 6 meV \cite{Stebe77,Stebe78}. CuCl also
has not typical structure of the top of the valance band (the top
subband is twofold degenerate). This in principle makes possible
to obtain electronic spin polarization close to 100\% by optical
pumping and without using additional magnetic field pulses. To
reach this goal we propose to illuminate the sample with
circularly polarized light of appropriate frequency, tuned to the
electron trion transition. During the illumination electrons with
only one defined spin projection on the direction of the light
propagation can participate in the trions creation, while the
electrons having opposite spin projection are not affected by
light. When the coherent light is used, the populations of trions
and participating electrons oscillate with Rabi frequency. Because
created trions can spontaneously decay to electrons with both spin
projections, the densities of electrons and trions participating
in oscillations diminish after some time in favor of rising
density of electrons non affected by light. This, in turn,
enhances the spin polarization.

In order to estimate the trion lifetime which influences the rate
of the spin pumping, we have used the effective mass approximation
to describe the trion bound state and for the calculation of
appropriate matrix elements for optical transitions. The methods
used, the results of the calculations and relevant selection rules
are given in section II. In the section III we explore the problem
of the pumping dynamics in the presence of the light pulse
propagating through the sample using the density matrix approach
and Maxwell equation. We consider charged sample with small amount
of excess free electrons, which are not generated by band to band
transitions. Such situation may, for instance, correspond to extra
carriers introduced by injection or by contact with n-doped
semiconductor and/or simultaneously applying electric field. We
assume that the temperature is close to zero and we deal with the
fully occupied valence band before and after pumping.

\section{\label{sec:level1}FREE ELECTRON $\leftrightarrow$ TRION $X^-$ TRANSITIONS}

CuCl has direct band gap $E_G$ = 3.43 eV at \textbf{k}=0 and
tetrahedral symmetry $T_d$ . The lowest conduction band of
symmetry $\Gamma_6$ and the uppermost valence band of symmetry
$\Gamma_7$ are both twofold degenerate at \textbf{k}=0. A lower
fourfold degenerate valence band of symmetry $\Gamma_8$  is
split-off   by the spin orbit interaction by
$\Delta(\Gamma_7-\Gamma_8)$= 69 meV. The $\Gamma_6$ and $\Gamma_7$
Bloch states correspond to the angular momentum $J=1/2$.
Expressing the electron wave functions with the symmetry of $s,
p_x, p_y$    and $p_z$ orbitals as $S, X, Y$  and  $Z$
respectively, the conduction Bloch functions $\Gamma_6$ can be
written as
\begin{equation}
c_{1/2}=S|\alpha\rangle, \;\;\; c_{-1/2}=S|\beta\rangle
\end{equation}
and the valence Bloch functions are of the form
$$
v_{1/2}=\frac{1}{\sqrt{3}}[(X+iY)|\beta\rangle+Z|\alpha\rangle],
$$
\begin{equation}
v_{-1/2}=\frac{1}{\sqrt{3}}[(X-iY)|\alpha\rangle-Z|\beta\rangle],
\end{equation}
where the indexes $+1/2$ and $-1/2$ denote projections $m_j$ of
the total angular momentum of  band electron onto $z$ axis, while
$|\alpha\rangle$ and $|\beta\rangle$ denote the parallel and
antiparallel to \emph{z} projections of pure electron spin.

The trion $X^-$ is formed by the three quasi particles, two
electrons and one hole, interacting with the Coulomb field.  In
absence of an applied magnetic field the negatively charged
exciton has only one bound state corresponding to the two electron
spin singlet \cite{Hill}. Then we assume the bound state of $X^-$
to be a linear combination
$$
|X^-;\textbf{ K},
M_j=\pm1/2\rangle=\frac{1}{\sqrt{2}}\sum_{\textbf{k}_1,\textbf{k}_2,
\textbf{k}_h}C(\textbf{k}_1, \textbf{k}_2,\textbf{k}_h)\times
$$
\begin{equation}
(a^+_{\textbf{k}_1,+1/2}a^+_{\textbf{k}_2,-1/2}
-a^+_{\textbf{k}_1,-1/2}a^+_{\textbf{k}_2,+1/2})d^+_{\textbf{k}_h,\pm1/2}|g\rangle\;\;,
\end{equation}
where \textbf{K} is the total momentum of $X^-$, $|g\rangle$
denotes the electronic state corresponding to the empty conduction
and fully occupied valence band, and $ a^+_{\textbf{k},m_j}$
($d^+_{\textbf{k},m_j}$) denote the creation operator of an
electron (hole) in the Bloch states with the wave vector
$\textbf{k}$ and projection of the angular momentum  $m_j$. In the
effective mass approximation the linear coefficients
$C(\textbf{k}_1, \textbf{k}_2,\textbf{k}_h)$, subjected to
$\textbf{k}_1+\textbf{k}_2+\textbf{k}_h=\textbf{K}$, are the
Fourier transforms of the trion $X^-$ envelope
$\Phi(\textbf{r}_1,\textbf{r}_2,\textbf{r}_h)\sim
exp(i\textbf{KR}_0) \psi(\textbf{r}_{1h},\textbf{r}_{2h})$ , where
$\textbf{R}_0$ denotes the center of mass vector of the whole
complex $X^-$, while $\textbf{r}_{1h}$, $\textbf{r}_{2h}$ are the
relative coordinates of the two electrons (1, 2) and one
hole($h$).

In order to investigate the free electron - trion transitions we
calculate the appropriate dipole matrix elements in a similar way
as given by Stebe et al \cite{Stebe98}. For the electric dipole
transition between the trion state with $M_j=+1/2$ and the free
electron states $a^+_{\textbf{k},m_j}|g\rangle$ with
$m_j=\pm{1/2}$ we get
$$
\langle g|
a_{\textbf{k},m_j}exp(-i\textbf{qr})\hat{\textbf{p}}|X^-;\textbf{K},+1/2\rangle=
$$
\begin{equation}
 \frac{1}{\sqrt{3}}\langle Z|\hat{\textbf{p}}|S\rangle
 I(\textbf{k},\textbf{q})\delta_{\textbf{K},
 \textbf{k}+\textbf{q}}
\end{equation}
 for $m_j$=+1/2,  and
\begin{equation}\frac{1}{\sqrt{3}}\langle X-iY|\hat{\textbf{p}}|S\rangle
I(\textbf{k},\textbf{q})\delta_{\textbf{K}, \textbf{k}+\textbf{q}}
\end{equation}
for $m_j=-1/2$. The factor $I(\textbf{k}, \textbf{q})$ depends on
the electron and emitted photon wave vectors $\textbf{k}$ and
$\textbf{q}$, respectively, and is defined as the integral
$$
I(\textbf{k}, \textbf{q})=
\sum_{\textbf{l}}C(\textbf{k},\textbf{q}+\textbf{l},\textbf{l})=
$$
\begin{equation}
\int_Vexp[-i\mu\textbf{kr}+(1-\mu)\textbf{qr}]\psi(\textbf{r},0)d^3r\;\;\;
,
\end{equation}
with $\mu=(1+\sigma)/(1+2\sigma)$ (the formula (4) coincides with
that given by Stebe et al \cite{Stebe98} when $\textbf{q}=0$).

Due to the selection rules of the dipole transitions resulting
from the Eqs. (4) and (5), there are two channels of spontaneous
decay of $X^-_\uparrow$: first, corresponding to the transition
$X^-_\uparrow \rightarrow e_\uparrow$ with the rate $w_1$, and the
second one, corresponding to the transition $X^-_\uparrow
\rightarrow e_\downarrow$ with the rate $w_2$ (hereafter for the
trion $X^-$, as well as for the electron $e$ we use the arrows
$\downarrow$ and $\uparrow$ to denote the angular momentum
projections $-1/2$ and $+1/2$ onto $z$ axis). Obviously, the same
rates correspond to $X^-_\downarrow$ - decay, i.e. $w_1$ for
$X^-_\downarrow \rightarrow e_\downarrow$ and $w_2$ for
$X^-_\downarrow \rightarrow e_\uparrow$. Due to $T_d$ symmetry of
CuCl, the matrix elements $\langle X|\hat{p}_x|S\rangle, \langle
Y|\hat{p}_y|S\rangle, \langle Z|\hat{p}_z|S\rangle$ are all equal
to each other and from Eqs. (4) and (5) it follows that the ratio
of rates is $w_1/w_2=1/2$.

In order to estimate the values of the rates $w_{1,2}$, we have
calculated the integral $I$ with the envelope
$\psi(\textbf{r}_{1h},\textbf{r}_{2h})$ approximated by the
calibrated wave function of the $H^-$  ion, where the wave
function of $H^-$ was taken as proposed by Rotenberg and Stein
\cite{Rotenberg}. Such an approximation can be justified by the
very small value of $\sigma=0.02$ in CuCl. The obtained value of
$I^{2}$ (called the envelope oscillator strength) decreases much
slower with $k$, than $I^{2}$ obtained by Stebe et al
\cite{Stebe98} for $\sigma=0.1$ or $\sigma=1$. At \emph{k }=0 we
have obtained $I^{2}=676$, which is very close to the value
obtained for small $\sigma$ in ref. \cite{Stebe98}. Using the
experimental data: $|\langle X|\hat{p}_x|S\rangle|^2/m\approx 3$
eV, $\hbar\omega\approx 3.2$ eV (energy of emitted photon),
$\eta\approx 1.9$ (the refractive index) and integrating over all
photon directions we estimate that the rate

\begin{equation}
w_1\approx\frac{4e^2\omega\eta}{9\hbar c^3m^2}|\langle
X|\hat{p}_x|S\rangle I|^2\approx1.2\cdot10^{11}s^{-1} \;\;\;,
\end{equation}

and the radiative lifetime of the trion
$\tau=(w_{1}+w_2)^{-1}\approx$ 2.8 ps.

In the case of resonant transitions induced by coherent
illumination, the electron and trion densities oscillate with Rabi
frequency
\begin{equation}\label{eqOmega}
\Omega=2dE/\hbar \;\;\;,
\end{equation}
where $E$ is the electric field modulus and $d$ denotes the
magnitude of the dipole moment of the electron - trion transition
$d=\sqrt{2/3}|\langle X|\hat{p}_x|S\rangle|Ie/m\omega$.

\section{\label{sec:level1}SPIN PUMPING}

Let us consider as a pumping light the $\sigma^+$ circularly
polarized coherent plane wave pulse with frequency $\omega$ and
electrical vector
$\textbf{E}(z,t)=\sqrt{2}E(z,t)[\hat{\textbf{x}}cos(\omega
t-qz)+\hat{\textbf{y}} sin(\omega t-qz)]$,  entering and
propagating in a sample along $z$ direction. The shape of the
envelope $E$ can change in time due to the coherent coupling of
the light with electrons and trions. For the $\sigma^+$ polarized
light transmitted along the $z$ axis the only possible induced
transitions are the transitions between electron states with
$m_j=-1/2$ and the trion states with $M_j=+1/2$. The $\sigma^+$
signal does not affect the electrons with $m_j=+1/2$, because the
trion $X^-$ with $M_j=3/2$ does not exist. Trion $X^-$ has a short
lifetime and can recombine spontaneously into the free electron
states. As it is described in the previous section the trion $X^-$
with $M_j=+1/2$ decays spontaneously into the electron state with
either $m_j=+1/2$ or $m_j=-1/2$, with the relevant rates
$w_1=1/3\tau$ and $w_2=2/3\tau$, respectively. So under the
influence of $\sigma^+$ polarized light the density of electrons
with $m_j=+1/2$ increases, while density of electrons with
$m_j=-1/2$ decreases, with analogy to the standard optical pumping
of electron spins in atoms, which is based on the depletion of one
of the ground-state sublevels and accumulation in the other spin
sublevel \cite{Brossel}. The $w_1/w_2$ ratio is given by the
symmetry properties of the wave functions. We should emphasize
however that the pumping process described here can succeed for
any $w_1/w_2$ providing it is not to small. If bigger is $w_1$ the
faster is the pumping rate.

\begin{figure}\centering
\includegraphics[width=0.5\columnwidth]{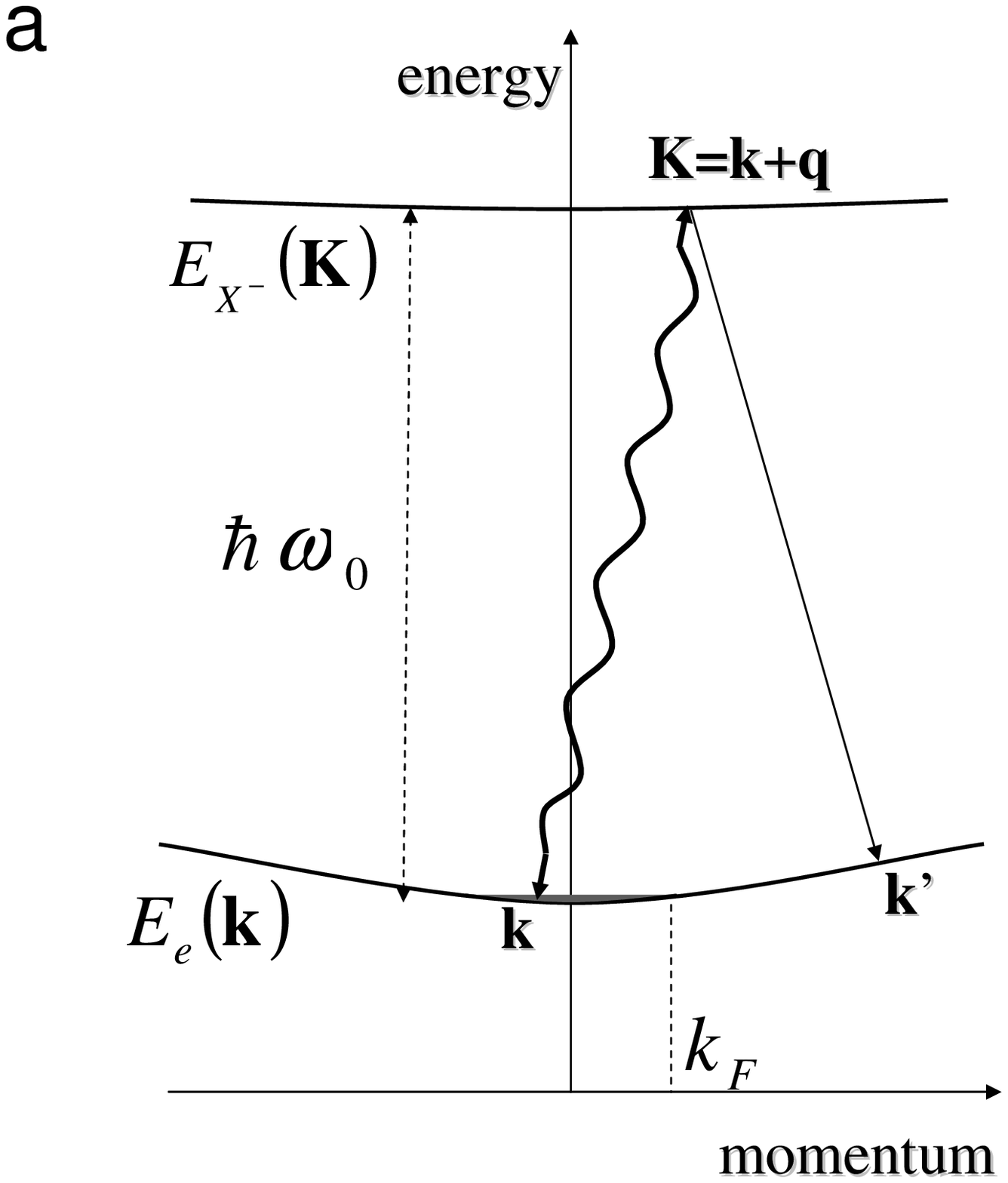}%
\hfill%
\includegraphics[width=0.5\columnwidth]{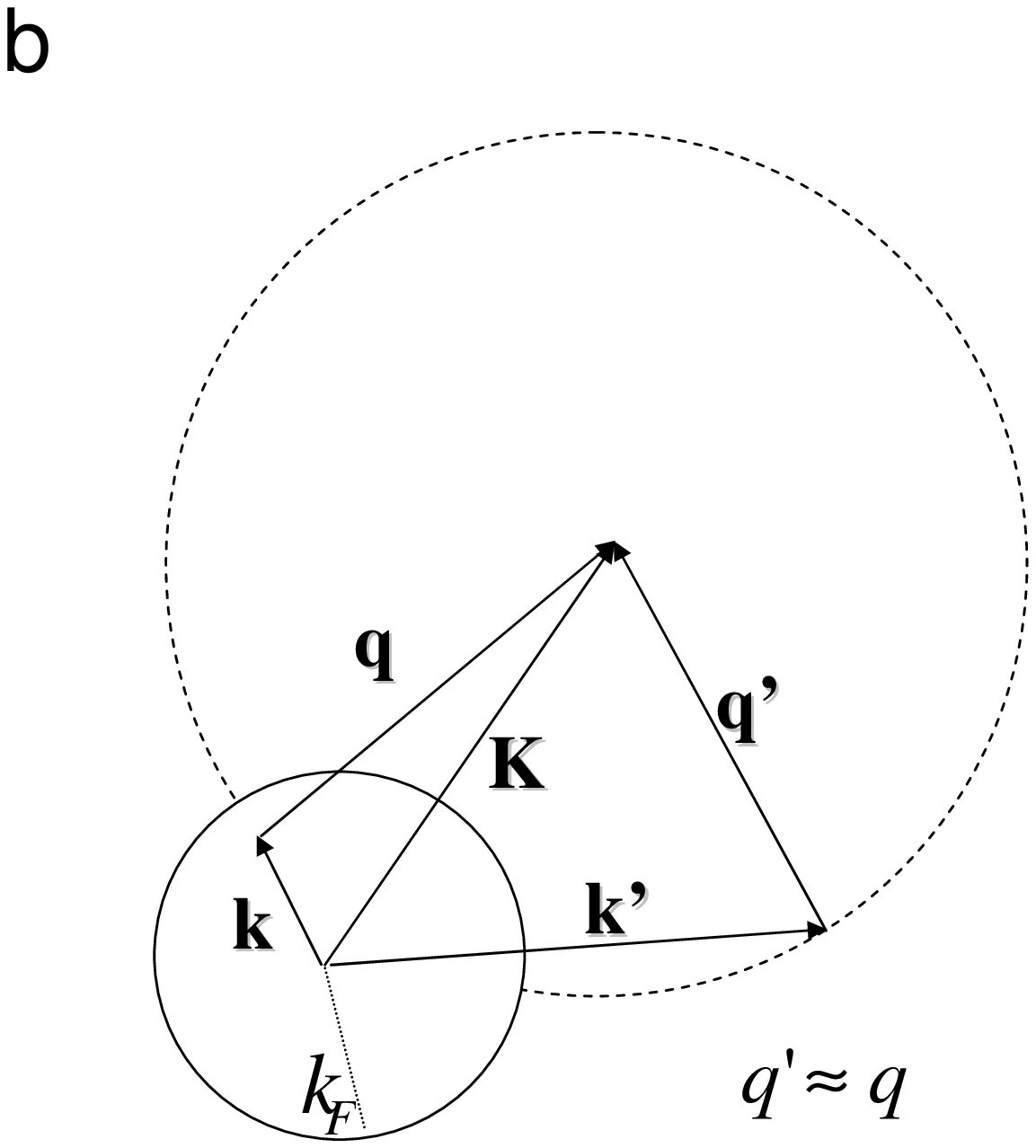}
\caption{\label{const} (a) Induced (wavy arrows) and spontaneous
(solid arrow) transitions between conduction electron and trion
$X^-$. (b) The possible electron states (dots) to which trions can
recombine spontaneously, defined by the momentum
($\textbf{K}=\textbf{k}'+\textbf{q}'$) and energy
($E_{X^-}(\textbf{K})=\hbar cq'+E_e(\textbf{k}')$) conservation.}
\end{figure}

During the decay of trions the photon transition energies fulfill
the equation $\hbar\omega=E_{X^-}(\textbf{K})-E_e(\textbf{k})$\;,
where $E_e(\textbf{k})=\hbar^2\textbf{k}^2/2m_e$ is the energy of
electron in conduction band, $E_{X^-}(\textbf{K})
=E_{X^-}+\hbar\textbf{K}^2/2(2m_e+m_h)$ is the energy of the trion
with $\textbf{K}=\textbf{k}+\textbf{q}$ and
$E_{X^-}=E_G-E_{b(ex)}-E_{b2}$ ($E_{b(ex)}$ denotes binding energy
of a free exciton). The photon energies lie below the threshold
$\hbar\omega_0\approx E_{X^-}$ (see Fig.1a) and correspond to the
photon wavelength $\lambda\approx$ 384nm. If the Fermi momentum
$k_F$ is sufficiently small in comparison to the photon wave
number q (i.e the Fermi energy of the excess electrons is
respectively smaller than
$\hbar^2q^2/2m_e=(\hbar\omega)^2\eta^2/2m_ec^2\approx$ 0.09 meV,
what corresponds to the concentration of electrons $N < N_0 =
10^{15}$ cm$^{-3}$) then practically all electron states to which
trions can recombine, are available (see Fig. 1b). Moreover, if
the spectral width of the pumping light signal $\Delta\hbar\omega$
is close to the trion linewidth ($\approx 0.22$ meV),
corresponding to its short lifetime, than all photons in the
pumping pulse can participate in the pumping process. Since the
oscillator strength $I^2$ does not change significantly with $k$
up to a few values of $q$, it follows that almost all electron
states, to which trions recombine spontaneously, can be involved
again in the process of spin pumping with similar probability. It
can happen that after multiple processes of induced transitions
and spontaneous decays some electrons can be moved in the
Brilloune zone far from its center and reach the places where they
do not participate any longer in the process while still being in
the state $m_j = -1/2$. However, we can safely assume that the
probability of such an event to occur in the time scale of few
tens of ps, which is typical time scale for electron energy
relaxation processes in semiconductors, is very small. In the
consequence, we can assume that all electrons with spin down can
participate in the spin pumping.

\begin{figure}\centering
\includegraphics[width=0.3\textwidth]{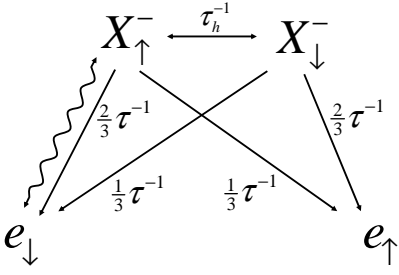}
\caption{\label{const}Scheme of transitions between electron and
trion states. Wavy line denotes the only possible transitions
which are induced by $\sigma^+$ circularly polarized laser light.
Incoherent transitions are denoted by the arrows accompanied by
the rates of the spontaneous decay of $X^-$.}
\end{figure}

Besides the induced coherent Rabi oscillations and incoherent
trion decays also the incoherent interaction of electrons and
trions with the environment should be taken into account and we
describe them in our model with the phenomenological parameters.
In our model we do not take into account the mobility of the free
electrons and trions, and what follows, the diffusion of
electronic spin, since it can be neglected in the timescales given
by speed of light as well as by trionic spin lifetimes. We assume
the long spin lifetime of an electron at (or near) the bottom of
the conduction band, however we take into account the possible
spin flips of trions. The total spin of trion is determined by the
spin of hole. We expect that because of the strong spin-orbit
interaction for electrons in the valence band the hole spin
lifetime $\tau_h$ in CuCl is short and of the order of few
picoseconds (We do not know the exact hole spin relaxation time in
CuCl, but we assume that it is of the order of typical time in
cubic semiconductors). We expect the same for the spin lifetime of
the trion, and assume that it is equal to $\tau_h$. Taking into
account the possible trion spin flips, we consider the system of
four (two electron and two trion) states. The scheme of all the
possible transitions between these states is shown in Fig. 2. We
would like to note here, that the trion spin flips do not
interfere with electron spin pumping but, on the contrary, they
open the second competing channel for the pumping process. What's
more, because the state $|X^-_\downarrow\rangle$ decays
spontaneously into the electron state $|e_\uparrow\rangle$ twice
as fast as into the state $|e_\downarrow\rangle$, the spin flips
of the trions do not delay the electron spin pumping.

We describe the dynamics of the electron spin polarization with
the density matrix operator $\hat{\rho}$ depending on time and on
position in the sample and written in the basis of two electron
and two trion states. The time evolution equation for $\hat{\rho}$
in the rotated basis $|e_\uparrow\rangle$, $|e_\downarrow\rangle$,
$|X^-_\downarrow\rangle$, $|X^-_\uparrow\rangle$, with respect to
the laser light frequency $\omega$, is given below (the diagonal
elements of $\hat{\rho}$ are labelled with single index):
\begin{eqnarray}\label{eqro}
\nonumber
\dot{\rho}_{e_\uparrow}&=&\rho_{X^-_\uparrow}\cdot\frac{1}{3\tau}+
\rho_{X^-_\downarrow}\cdot\frac{2}{3\tau} \\ \nonumber
\dot{\rho}_{e_\downarrow}&=&-\Omega Im\rho_{e_\downarrow
X^-_\uparrow}+ \rho_{X^-_\uparrow}\cdot\frac{2}{3\tau}+
\rho_{X^-_\downarrow}\cdot\frac{1}{3\tau} \\
\dot{\rho}_{X^-_\uparrow}&=&\Omega Im\rho_{e_\downarrow
X^-_\uparrow}- \rho_{X^-_\uparrow}\cdot\frac{1}{\tau}-
(\rho_{X^-_\uparrow}-\rho_{X^-_\downarrow})\cdot\delta
\nonumber\\
\nonumber
\dot{\rho}_{X^-_\downarrow}&=&-\rho_{X^-_\downarrow}\cdot\frac{1}{\tau}+
(\rho_{X^-_\uparrow}-\rho_{X^-_\downarrow})\cdot\delta \\
\dot{\rho}_{e_\downarrow X^-_\uparrow}&=& \frac{i}{2}\Omega
(\rho_{X^-_\uparrow}-\rho_{X^-_\downarrow})- \rho_{e_\downarrow
X^-_\uparrow}\cdot\gamma\;\;,
\end{eqnarray}
where $ \gamma$ is the transverse dumping rate and
$\delta=1/\tau_h$. The time evolution equation (\ref{eqro}) is
coupled to the Maxwell equation for the electric field by the
relation
\begin{equation}\label{eqE}
\frac{\partial E}{\partial z}+\frac{1}{v}\frac{\partial
E}{\partial t}= \frac{2\pi qNd}{\eta^2}Im\rho_{e_\downarrow
X^-_\uparrow} \;\;,
\end{equation}
where envelope $E$ is a slowly varying function of $z$ and $t$ and
the right hand side of Eq.(\ref{eqE}) corresponds to the extra
polarization due to the induced transitions between the free
electron and trion states \cite{McCall,Allen}. The refractive
index $\eta$ (and the velocity of light in the sample $v=c/\eta$)
is assumed to be unaffected by the light intensity, as the plasma
frequency for considered free electron concentrations
$\omega_p=\sqrt{4\pi Ne^2 /m}$ is lower than $10^{-4}\omega$.

\begin{figure}\centering
\includegraphics[width=\columnwidth]{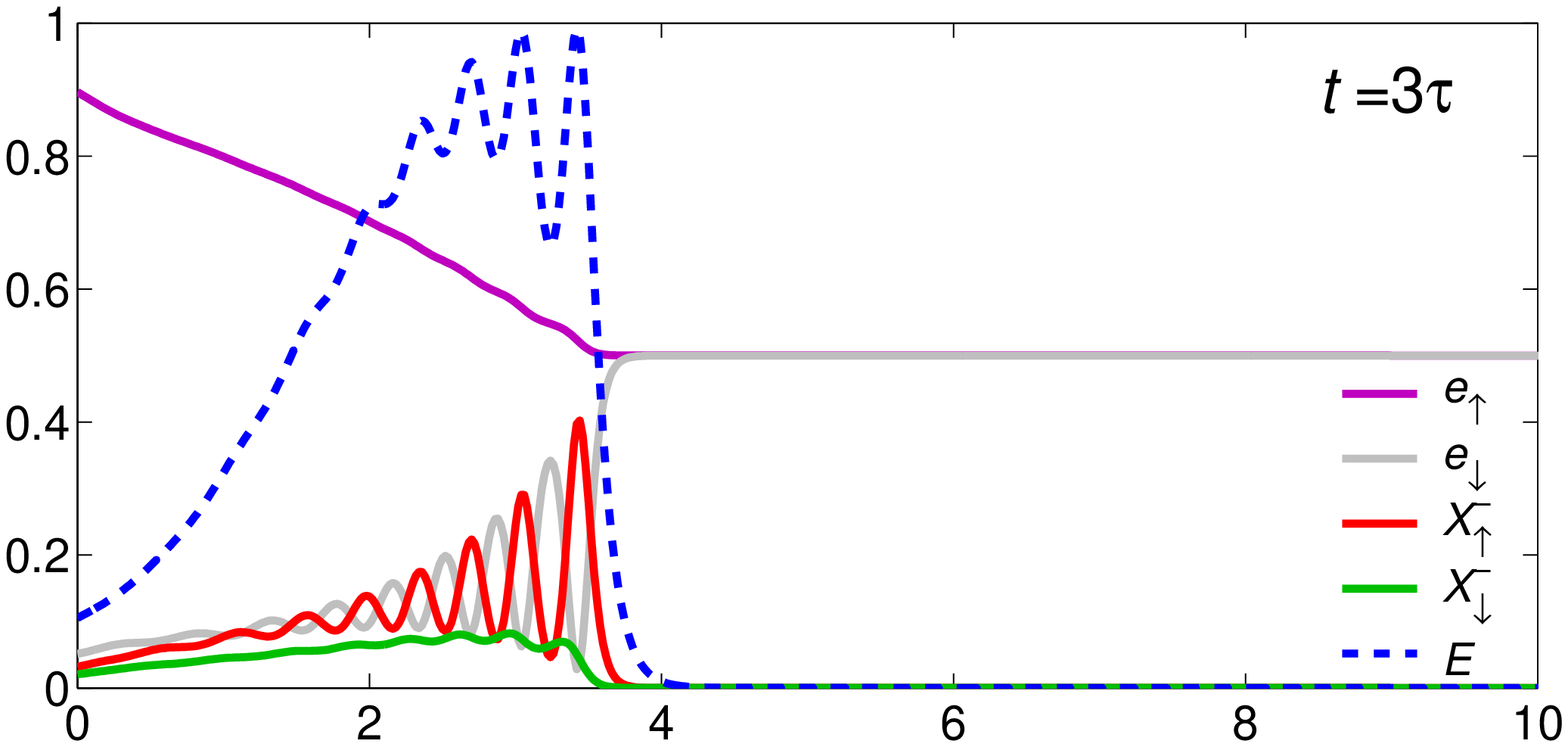}
\includegraphics[width=\columnwidth]{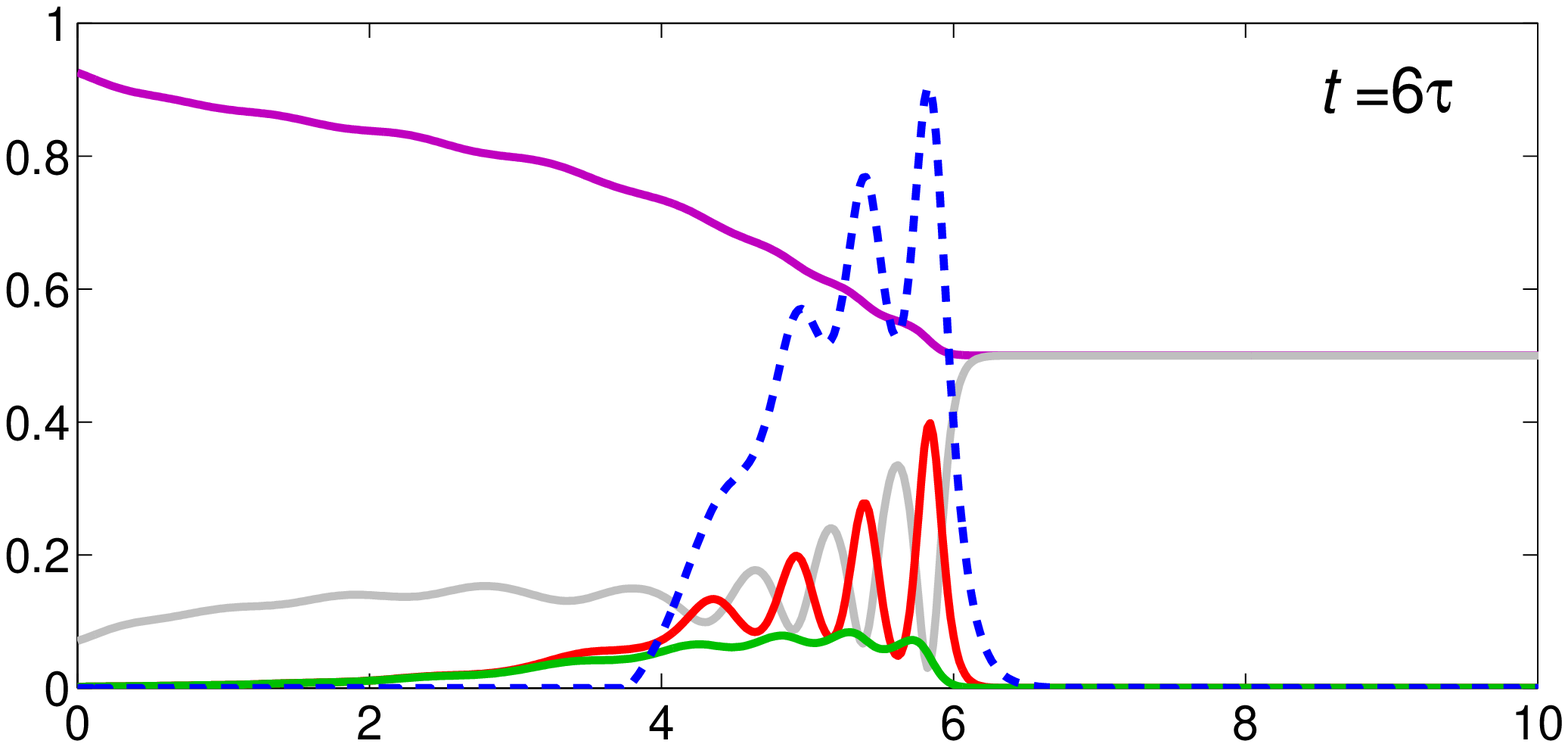}
\includegraphics[width=\columnwidth]{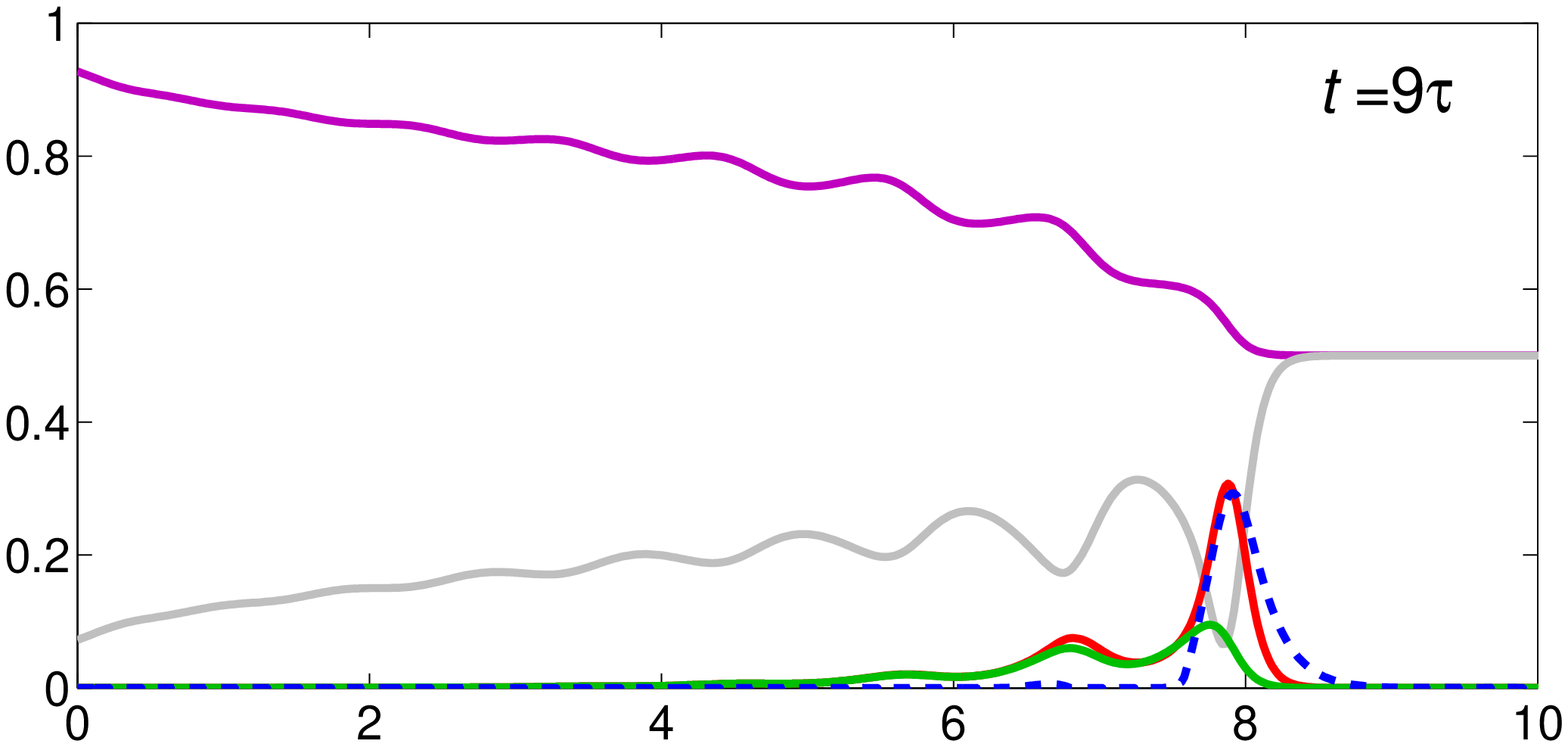}
\includegraphics[width=\columnwidth]{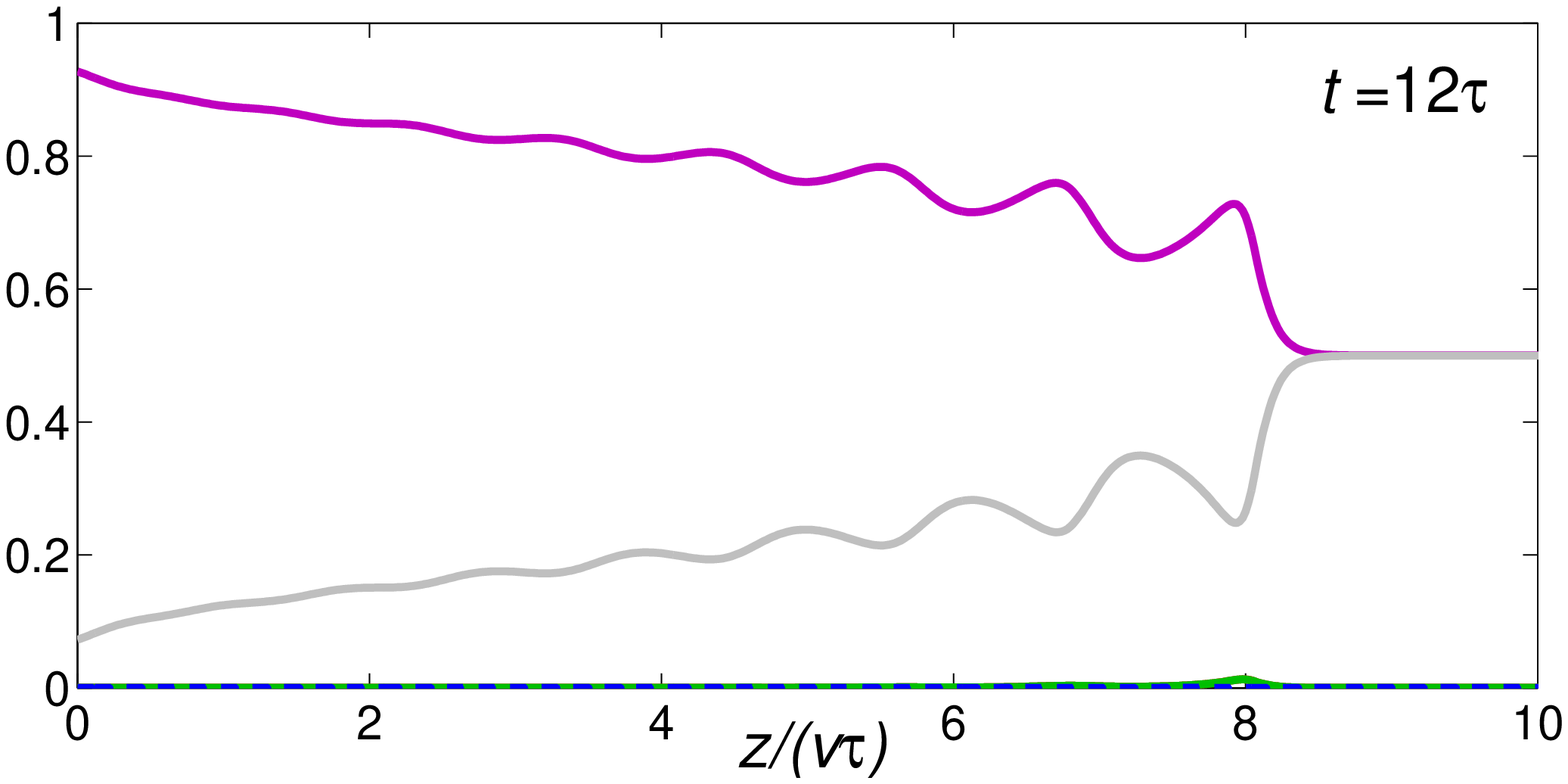}
\caption{\label{const}(color online) Four snapshots show the
effects of the Gausian resonant light pulse on the electronic and
trionic populations at different time moments. The electric field
envelope $E$ is drawn with the blue curve and is normalized to its
maximum value in $t=0$. The initial width of the pulse before
entering to the sample is equal to $2\tau$. Other color curves
correspond to the densities of $e_\uparrow$, $e_\downarrow$,
$X^-_\uparrow$, $X^-_\downarrow$ normalized to 1 (density matrix
diagonal elements). $z$ is the distance from the surface of the
sample. The results are given for the electron concentration $N
=2\cdot 10^{13}cm^{-3}$, maximum electric field envelope
corresponding to $\Omega_0\tau=16$ and $\gamma$ and $\delta$ rates
equal to $\tau^{-1}$.}
\end{figure}

We consider a CuCl sample with the small volume electric charge.
Let us imagine for example, that the surface plane $XZ$ of the
sample is in contact with appropriately chosen n-doped
semiconductor. It provides electrons that can be additionally
driven into CuCl crystal by the applied voltage in the $y$
direction. We assume then, that in the thin layer, parallel to the
$XZ$ plane, at the certain fixed distance from the contact, the
electron charge distribution can be treated as constant. We relate
considered above spin pumping process to such a layer. In the
practical realization of the proposed experiment it would be
important to take into account the real charge density
distribution dependent on the way the electrons are injected and
the geometry of the sample. Also the effects of the
electromagnetic wave scattering at sample boundaries and
interfaces should be taken into account. In our first attempt to
describe the process we neglect those effects and restrict our
study to the general properties of the model.

\section{\label{sec:level1}RESULTS AND DISCUSSION}

We assume, that the initial density of trions is equal to 0,
initial electronic spin densities ($n_\uparrow$) and
($n_\downarrow$) are both equal to $N/2$ and electrons are in the
mixed state (${\rho}_{e_\uparrow}={\rho}_{e_\downarrow}=1/2$). In
Fig.3 and 4 we present the solutions of the Eqs.
(\ref{eqOmega}),(\ref{eqro}) and (\ref{eqE}) for two types of
illumination and various input data. Fig.3 presents the influence
of the passing Gaussian light pulse on the electronic and trionic
spin populations (the appropriate diagonal density matrix
elements) for the chosen time values measured since the moment
when maximum value $E_0$ of the envelope $E$ enters into the
sample. It can be observed that during the illumination the Rabi
oscillations between $e_\downarrow$ and $X^{-}_\uparrow$ are
induced with the frequency $\Omega$ dependent on the temporary
value of $E$, as given in Eq. (\ref{eqOmega}). The energy stored
in the pulse is continuously dissipated due to the spontaneous
recombination of trions and finally the light pulse dies at
certain depth. The energy dissipation causes also the lowering of
the speed at which the pulse travels through the sample. The
results shown in Fig.3 present the case when the appropriately
short light pulse, after passing the sample, leaves behind the
nonmonotonic spin polarization. This effect is related to the
nonmonotonic dependence on $z$ of the final local Rabi oscillation
phase.

\begin{figure}\centering
\includegraphics[width=\columnwidth]{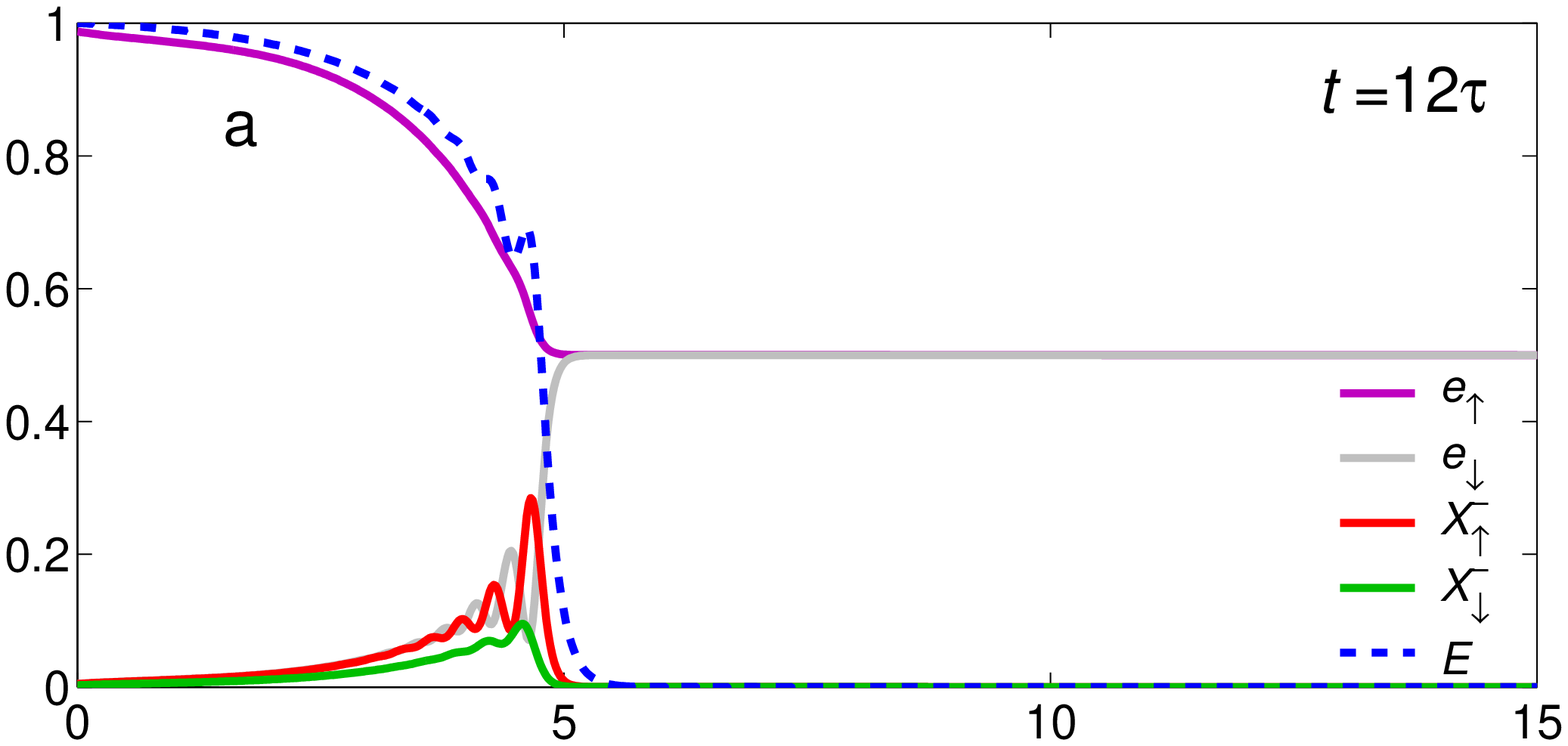}
\includegraphics[width=\columnwidth]{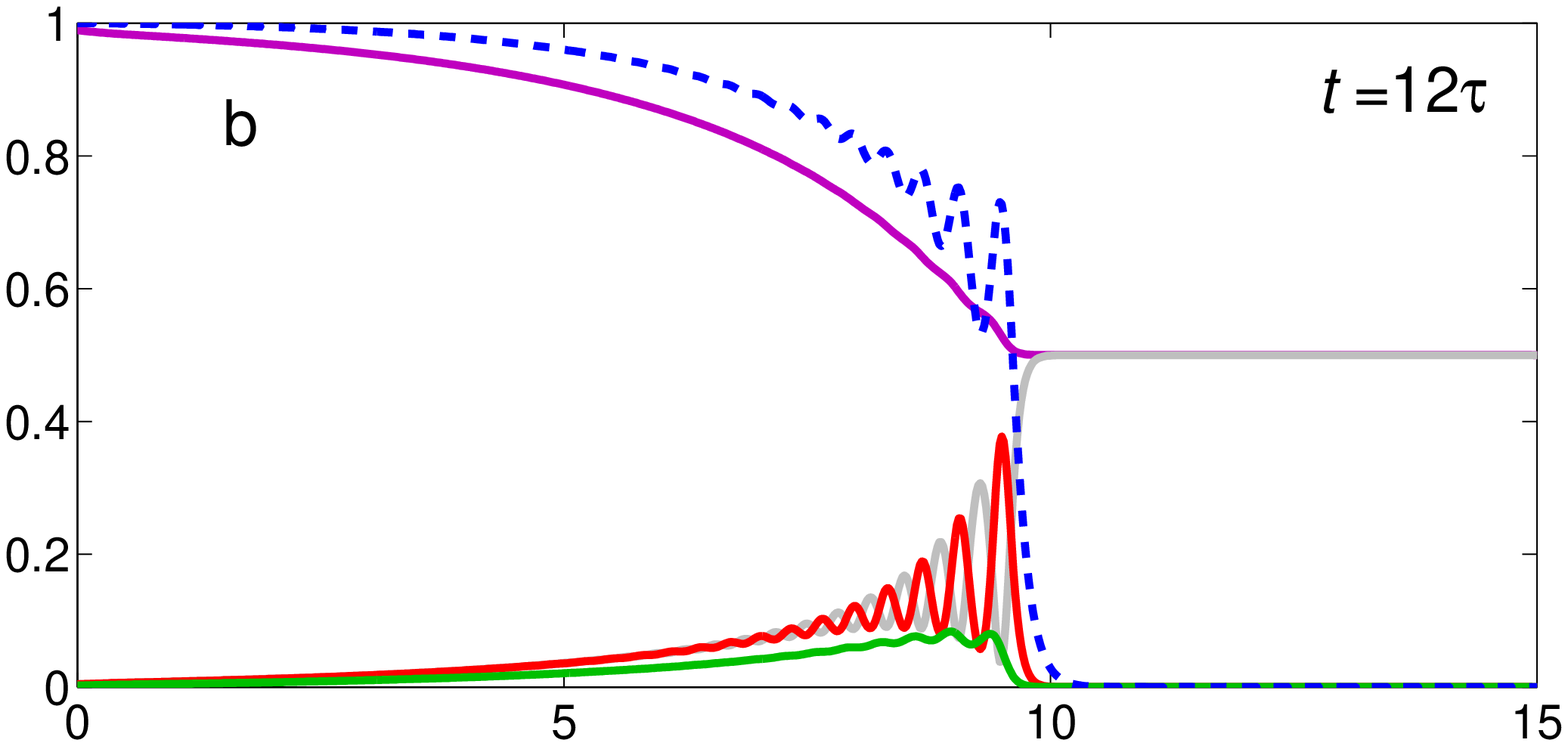}
\includegraphics[width=\columnwidth]{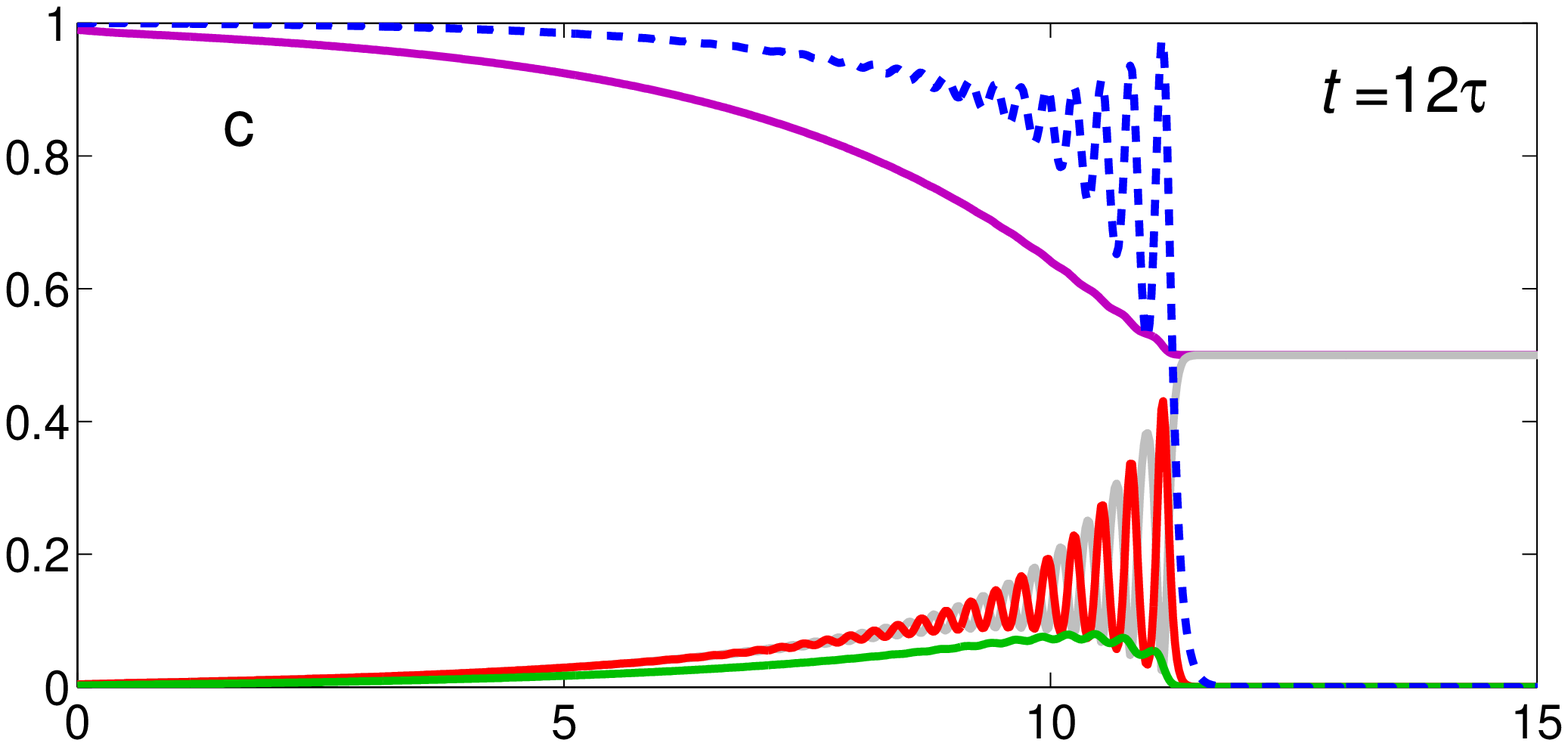}
\includegraphics[width=\columnwidth]{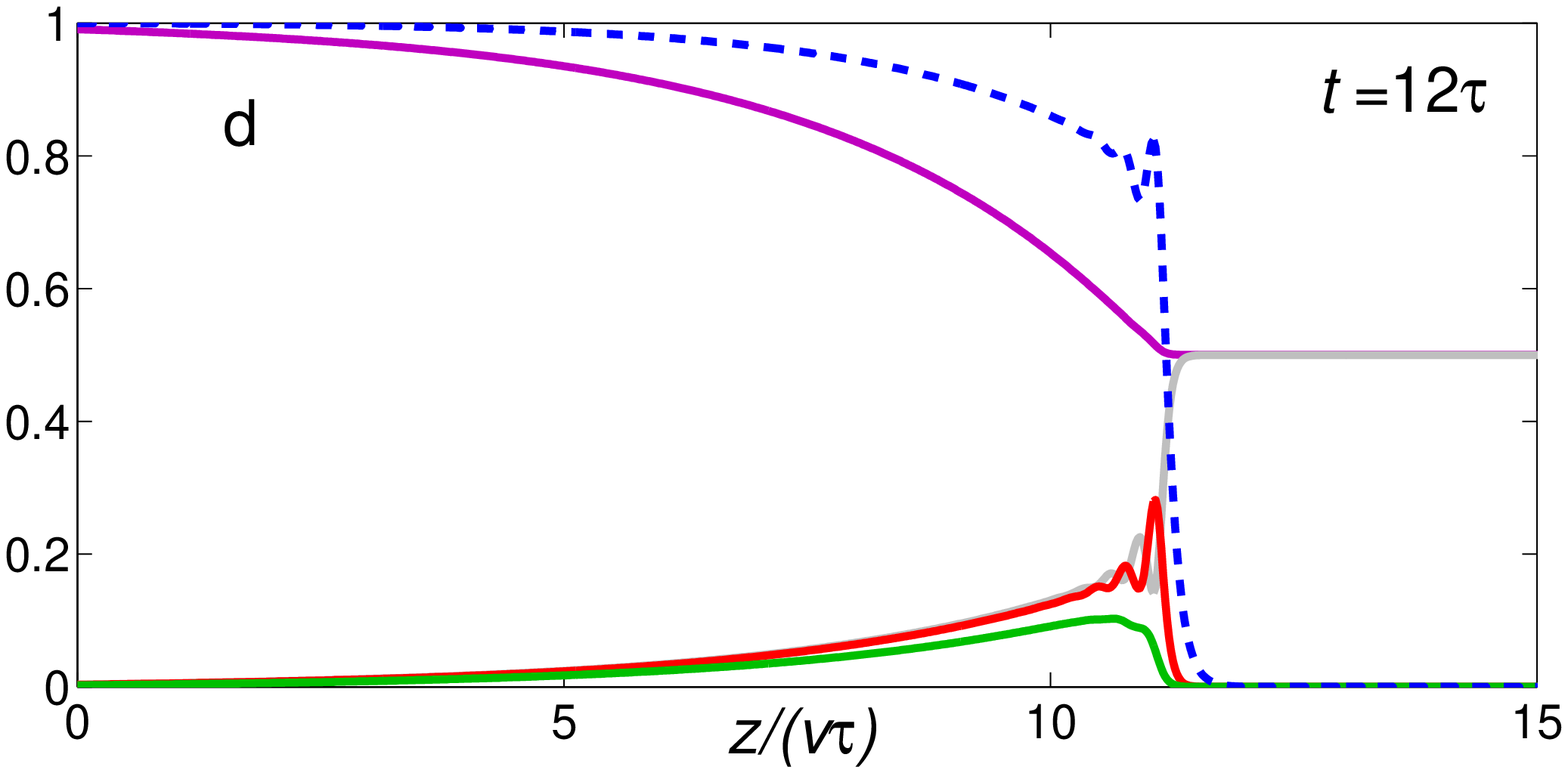}
\caption{\label{peak}(color online) The presented electronic and
trionic populations are obtained with the same data as the results
shown in the Fig.3 but the illumination is now constant and
corresponds to $\Omega_0\tau =8$, $16$, $24$ in the a, b and c
case respectively and $t=12\tau$ for all four cases. The results
in the d case have been obtained for the same illumination as in
the case c but with the higher rates $\gamma =8\tau^{-1}$ and
$\delta =2\tau^{-1}$. }
\end{figure}

Fig.4 shows four examples of the influence of the step-like shaped
signal with $E_0$ amplitude on the electron and trion spin
distributions (the step-like shape has been in fact kindly
smoothed out within the narrow width $v\tau$). The signal enters
the sample at $t=0$. After few trion lifetimes the front of the
travelling signal establishes its shape and all cases shown in
Fig.4 relate to time $t=12\tau$ when the front shape is already
constant. The same is true for the shapes of the spin populations.
In the case of small dumping rates, Rabi oscillations can be
observed just before the signal's front. The light pumps
electronic spins and makes the sample transparent. After
examination of many results, obtained with various parameters, we
have made some general observations. First, the speed of the
signal front depends on the intensity $E_0$ and on the electron
concentration $N$, and can be much smaller than the speed of light
in the crystal $v$. Below some critical value of $E_0^{2}/N$, the
front speed decreases with rising $N$ and with reduction of the
energy ($\sim E_0^2$) stored in the pulse (compare the position of
the front in the Fig. 4c with its positions in Fig. 4a and b).
Other observations are related to the speed of spin pumping. Since
the moment when the signal front reaches the given position $z$,
the pumping speed is determined mainly by the trion lifetime
$\tau$, provided that frequency of Rabi oscillations $\Omega >
\tau^{-1}$. The speed of pumping is also not noticeably influenced
by damping rates $\delta$ and $\gamma$, if their values are not
higher than 10$\tau^{-1}$. Nevertheless they influence the damping
rate of Rabi oscillations (compare Fig.4c with Fig.4d).

The presented results of the simulations of spin pumping dynamics
are obtained for $\Omega\tau\leq16$, corresponding to the density
of photons in the incident laser beam $E^2/2\pi\hbar\omega <
10^{13}cm^{-3}$, i.e. to the laser intensity radiation less than
150kW/$cm^2$, and situates the intensity within the range used in
the optical solid state experiments.

In our model we have not accounted for the depolarization of
electron spins caused by the photons spontaneously reradiated by
trions. This process can interfere with spin pumping and in
general is hard to estimate. However, the density of the photons
emitted spontaneously by trions and interacting with electrons is
limited by the size of the space where electrons are located. For
example, if the electrons are confined in the 2dim layer of the
thickness $r$ and parallel to the XZ plane, then density of
photons is lower than $N\rho_{X^-_\uparrow}r/v\tau$, where
$N\rho_{X^-_\uparrow}/\tau$ is of the order of the density of the
trions which decay in a unit of time. The rate of reabsorption of
the photons in the process of recurrent creation of $X^-$ with
electrons having spin $\uparrow$ is proportional to the density of
photons as well as to the transverse lifetime $\gamma^{-1}$
\cite{Allen}, which does not exceed $\tau$. Taken the above into
account one can find that the depolarization of electronic spins
is negligible under the condition $Nr < 5\cdot10^9cm^{-2}$. Thus,
when the thickness of the sample or the thickness of the space
where electrons are located is $r\approx1\mu m$ the concentration
of the excess electrons should be $N < 5\cdot10^{13}cm^{-3}$. The
thickness $r$ should not be too small, in order to exclude the
quantum confinement effects.

We have also neglected the effect of spin depolarization caused by
the electron-trion collisions assuming, that the electron-trion
scattering rates, similarly as the electron-electron scattering
rates in semiconductor, are usually negligible in the case of
small electron concentration, considered in our model. The elastic
collisions between electrons do not change the electron spin
densities and therefore they do not influence the pumping process.
The same can be said about the elastic collisions between trions.

The separate remark should be made concerning the role of possible
creation of excitons by the pumping light pulse. Even if the
energy separation between trion and exciton is large, the high
density of valence electrons in comparison with the low
concentration of excess electrons can make the probability of
excitonic creation comparable to the probability of the creation
of trions. However, when excitons collide with electrons in
inelastic way, then the second channel opens for trions creation.
The examination of the appropriate selection rules makes clear,
that exciton created by $\sigma^+$ optical pulse can form trion
only with the electron in the state $|e_\downarrow\rangle$ , what
means, that the new channel does not interfere with the first
channel discussed here, and it can influence only the pumping
rate.

Our model predicts that once the light reaches the given place
then close to 100\% polarization of electronic spins associated
with full transparency of the crystal can be obtained there in the
time shorter than about $20\tau$. Assuming that the sample has a
size in the $z$ direction of the order of few mm and taking into
account the time needed for the front of the light pulse to cross
the sample, we can estimate, that in the case of high enough
illumination all electronic spins in the sample can be fully
polarized in the time scale of the order of 100ps.

In conclusion we propose the efficient and fast method of the
electron spin pumping induced by coherent laser light in the CuCl
crystal. We have used several approximations for the estimation of
the trion lifetime and for the description of the pumping
dynamics. We have obtained the timescale of the process and the
results showing interesting effects of pumping (like possible
spacial modulation of spin polarization and slow down of the speed
of the front of the light pulse). The incoherent interactions with
environment have been described by using phenomenological
parameters, which are not well known. We have examined the pumping
dynamics in the broad range of these parameters observing that
they influence the pumping rate very little, having however high
impact on the dumping of Rabi oscillation. The spin pumping rate
depends mainly on the $w_1$ rate (equal to $1/3$ of the inverted
trion lifetime). The proposed model can be used in the case of
very small density of free excess electrons, which can be
challenging to control in the experiment. In the possible
experimental realization many factors which can influence the
pumping dynamics, like real sample geometry, light scattering and
interference effects as well as inhomogeneous electron
concentration, should be taken into account. The degree of spin
polarization can be evaluated in the experiment by measuring for
example the transparency of the sample in the case of the
illumination with circularly polarized light. CuCl has inverted
(if compared to other typical semiconductors) valence band
structure and relatively large binding energy of the negatively
charged excitons. We expect, that similar electron spin
polarization through the intermediate trion states is also
possible, but probably not as effective, in other materials with
typical valence band structure and usually having also much lower
binding energy of trions. The efficiency of the process in these
materials should be higher in two dimensional structures, where
energy separation between trions and excitons is larger.

\begin{acknowledgments}
This work was supported by The Polish State Committee for
Scientific Research grant PBZ-KBN-044/P03/2001 and by Polish
Ministry of Scientific Research and Information Technology grant
PBZ-MIN-008/P03/2003.
\end{acknowledgments}

\end{document}